\documentclass[onecolumn]{emulateapj}
\newcommand{\msun}{M$_{\sun}$}
\newcommand{\kms}{km s$^{-1}$}
\newcommand{\msuns}{M$_{\sun}~$}
\newcommand{\kmss}{km s$^{-1}~$}

\begin{document}

\title{Capture Formed Binaries via Encounters with Massive Protostars }

\author{Nickolas Moeckel, John Bally} 
\smallskip
\affil{ Center for Astrophysics and Space Astronomy, and\\
        Department of Astrophysical and Planetary Sciences \\
	University of Colorado, Campus Box 389, Boulder, CO 80309-0389}
\email{moeckel@colorado.edu}

\begin{abstract}
Most massive stars are found in the center of dense clusters, and have a companion fraction much higher than their lower mass siblings; the massive stars of the Trapezium core in Orion have $\sim 1.5$ companions each.   This high multiplicity could be a consequence of formation via a capture scenario, or it could be due to fragmentation of the cores that form the massive stars.  During stellar formation circumstellar disks appear to be nearly ubiquitous.  Their large radii compared to stellar sizes increase the interaction radius significantly, suggesting that disk interactions with neighboring stars could assist in capturing binary companions.  This mechanism has been studied for stars of approximately solar mass and found to be inefficient.  In this paper we present simulations of interactions between a 22 \msuns star-disk system and less massive impactors, to study the disk-assisted capture formation of binaries in a regime suited to massive stars.  The formation of binaries by capture is found to be much more efficient for massive capturers.  We discuss the effects of a mass dependent velocity dispersion and mass segregation on the capture rates, and consider the long term survival of the resultant binaries in a dense cluster.
\end{abstract}


\section{INTRODUCTION}

The high multiplicity of massive stars \citep{mas98,pre99,sta00,gar01} may provide clues to their formation.  A useful number to describe the multiplicity  is the companion star fraction \citep{zin02}, $csf = (B + 2T + 3Q + ...)/(S + B + T + ...)$, with $S$ the number of single stars, $B$ the number of binaries, $T$ the number of triples, and $Q$ the number of quadruples.  In the Orion Nebula cluster, the Trapezium stars have $csf = 1.5$, high compared to $csf \sim 0.5$ for low mass stars in the cluster.  Fragmentation and capture are the two most likely mechanisms by which these binary and higher multiple systems form.  Capture could occur dynamically, via a three body encounter, or by dissipation of orbital energy through disk or envelope interactions.

Capture by a disk has been studied as a binary formation mechanism by previous authors.  \citet{cla91} performed an analytical approximation to derive an upper limit to the capture rate for roughly 1.0 \msuns star-disk systems, concluding that the rate is too low to contribute to the observed multiplicity.  \citet{hel95} used a smoothed particle hydrodynamics (SPH) code to study encounters between 1.0 \msuns stars, one of which is surrounded by a 100 AU, 0.1 \msuns disk.  These conditions, when the envelope is depleted and the prestellar material is found solely in a disk around the star, represents the class II phase of prestellar evolution.  He examined a range of inclination angles and periastra, and from these results determined a binary formation rate.  The rates found for this late stage of evolution are too low to implicate capture as a significant binary formation mechanism.

\citet{bof98} performed a similar analysis to Heller, but on a less evolved, more extended disk system.  The impactor was again 1.0 \msun, but the primary and disk were both 0.5 \msun, and the disk radius was 1000 AU, similar to a class 0 object.  The motivation for studying a less evolved system is the observation that by the time stars reach the class II and III stage of evolution, their binarity has been established \citep{mat94}.  Even with the increased cross sections of these larger disks, the capture rates in dense regions like the Orion Trapezium center are too low to account for the observed multiplicity fractions.

The early evolution of massive protostars is less well understood than lower mass cases. Disk or envelope capture during the formation of a massive star could make a significant contribution to their observed multiplicity.  Because of its stronger gravitational focusing, a 20 \msuns star with a 2 \msuns disk is a more attractive encounter partner for a lower mass cluster sibling.  When encounters occur, the interactions will involve a disk with a mass comparable to the impactor.  The lowest mass impactors will encounter a disk an order of magnitude more massive than themselves.  The regime with secondary stars significantly less massive than the disk has not been explored.

\citet{moe06} performed simulations of encounters between a massive star-disk system and a less massive intruder.  That work considered encounters at only one radius and impactor mass, focusing on the effects of repeated encounters on the disk.  In this paper we extend the analysis of encounters with massive protostars, and describe the results of simulations of encounters between intruder stars of several different masses and a 20.0 \msuns star surrounded by a 2.0 \msuns disk, 500 AU in radius.  Though the applicability of the standard prestellar classification scheme to massive protostars is unclear, this moderately-extended disk is similar to a later class II stage in the formation of a massive star.

There are several observational examples of disks around massive protostars, e.g. NGC 7538 IRS 1/2 \citep{kra06}, G192.16-3.82 \citep{she01}, Cepheus A HW2 \citep{pat05}, IRAS 20126+4104 \citep{ces99}, and MWC 349 \citep{whi85,taf04}.  A commonality among these sources is that they are all young, embedded protostars.  By the time a massive star has reached a more evolved stage, for instance in the Trapezium, disks are absent.  Either during or shortly after the embedded phase, disk destruction is apparently fast, via photoevaporation or other mechanisms.  The disk lifetime is then approximately the length of the embedded phase, $\sim 10^{5} - 10^{6}$ yr.  We show that over similar time periods, encounters with cluster siblings can contribute both to the formation of binaries and the destruction of the disks around massive stars.

\section{SIMULATIONS}
The code used in this study is a modified version of the publicly released SPH/N-body code GADGET-2 \citep{spr05}.  We model the stars as point masses interacting only through gravity, while the gas particles experience hydrodynamical forces as well.  We have modified the code so that the stars behave as sink particles \citep{bat95}, accreting gas that falls within the pre-set, mass dependent accretion radius.  The accretion radii are set to be small compared to the Bondi-Hoyle radius $r_{g} \sim 2GM_{\star}/(v^{2}+c_{s}^{2})^{1/2}$, where $c_{s}$ is the sound speed.  Using sink particles prevents the gas from clumping in unphysical manners about the stars, and prevents very dense gas from dominating the integration once it is very close to the stars.
  We have also modified the numerical viscosity slightly, to remain at low levels except when shocks are present \citep{mon97}.  For more detail on the code and the changes made to it, see \citet{moe06}.

We set up the disk with a surface density profile  
\begin{equation}
  \label{sigma}
  \Sigma(r) = \Sigma_{0} \left( \frac{r}{r_{0}} \right)^{-1},
\end{equation}
 
 with the vertical density structure at a given radius 
\begin{equation}
  \label{rho_init}
  \rho(r,z) = \rho_0(r) exp\left( -\frac{z^2}{2H(r)^2} \right),
\end{equation}
where $H(r)$ is a temperature dependent scale-height and $\rho_0$ the density at the mid-plane.  The smoothing lengths of the SPH particles are less than $H(r)/2$ for all but the innermost regions of the disk.

The temperature profile is
\begin{equation}
  \label{temp_profile}
  T(r) = T_{0} \left( \frac{r}{r_{0}} \right)^{-1/2},
\end{equation}
with T$_{0}$ set depending on the mass of nearest stellar particle.  To return to this temperature profile when the disk is perturbed, we implement a simple cooling scheme,
\begin{equation}
  \label{cooling}
  \frac{d u}{d t} = -\frac{u - u_{base}}{\tau_c}.
\end{equation}
Here $\tau_c$ is the cooling timescale, and is inversely proportional to the Keplerian orbital frequency at the particle's radius.  The temperature profile of equation (\ref{temp_profile}) is converted to an internal energy to provide $u_{base}$.  The disks are not allowed to cool below the base temperature profile; in the absence of this, the cooling timescale we use is short enough that fragmentation would likely occur \citep{gam01,ric05}, an effect we are not considering here.  The temperature and density profiles are appropriate to a disk shortly after its formation \citep{lin90}. 

\begin{figure}
 \centering
  \plotone{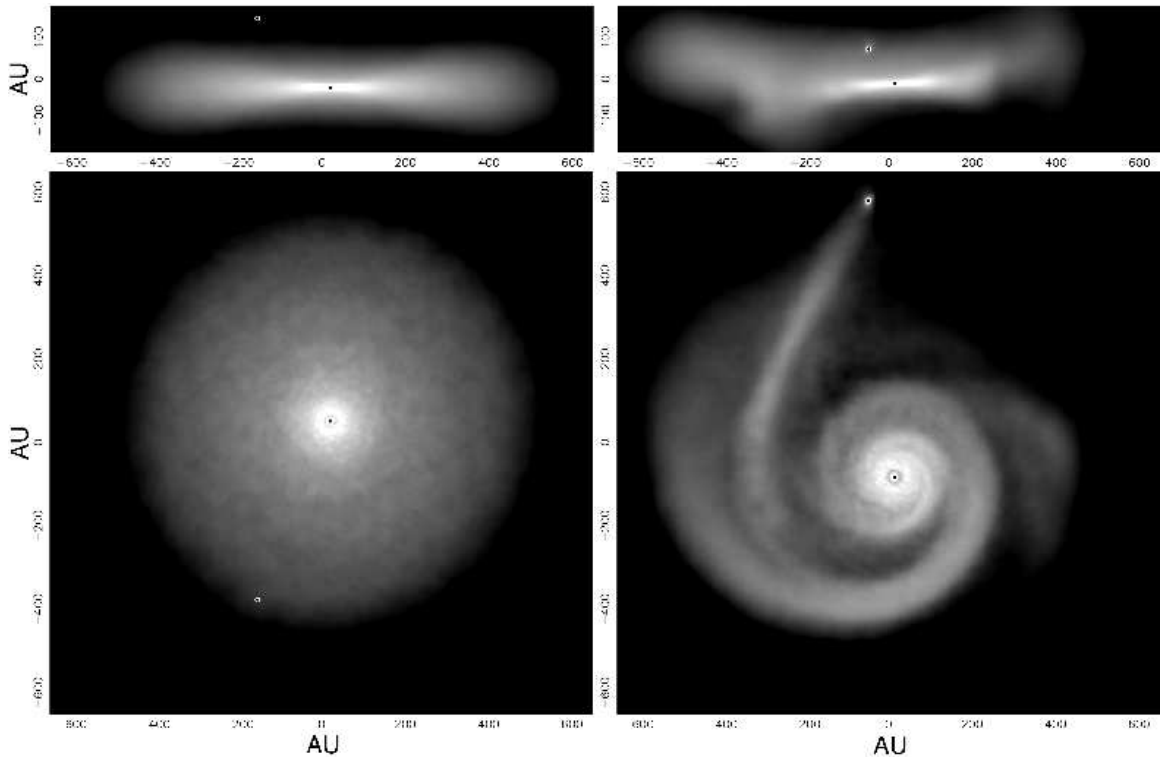}
  \caption{An example of an encounter, with $m = 3.0$ \msun, $r_{p} = 175$ AU, and $i = 45$ degrees.  Plotted is the logarithm of the gas column density.  {\it Left}: The impactor has not yet passed through the disk, which is in its relaxed, initial state. {\it Right}: The passage through the disk is completed, and spiral features are visible.}
  \label{encounter_example}
\end{figure}

We place $\sim 3.2\times10^{4}$ particles pairwise symmetrically around the central star, and normalize the surface density so that 2 \msuns is contained in the disk out to $r_{d}$ = 500 AU.  The temperature is normalized so that the disk is Toomre stable throughout.  We allow the disk to evolve in isolation until it has reached a steady state, at which point we introduce the impactor.  The initial disk is shown in the first panel of figure \ref{encounter_example}.  To test the resolution of our simulations, we ran a retrograde encounter with a 1 \msuns impactor, and periastron radius 0.6 times the disk radius, with $1.28 \times 10^{5}$ particles.  There was less than 3 percent difference in the orbital energy change between the lower and higher resolution simulations, with the lower resolution yielding a less tightly bound binary.  Since the lower resolution displayed less dissipation, the simulations performed here can be viewed as slightly conservative in the capture rate estimations.  If we were to allow fragmentation in the disk, a significantly higher number of particles would be required to capture the details.  Because the energy transfer in these simulations is dominated by the large scale behavior of spiral arms and trailing shocks, the relatively low resolution used is sufficient.

The impactor is initially at a distance of 1000 AU, on an orbit corresponding to a relative velocity at infinity of 2 \kms.  The simulations all have the same relative velocity, chosen to be representative of typical velocity dispersions in young clusters.  The eccentricities of the encounters that result in a bound system are all less than $\sim 1.1$; above this the velocities are too high for dissipation via disk interactions to create a bound system.  Tests suggest that the change in orbital energy is not a strong function of eccentricity for such low relative velocities.  We ran prograde and retrograde encounters with a 1 \msuns impactor using relative velocities of 0.1, 2, and 4 \kms, corresponding to eccentricities of 1.000, 1.058, and 1.235 respectively.  For the prograde case, the change in energy of the different cases agree to within 5 percent.  The retrograde case exhibits a slightly larger variance, approximately 10 percent, with the highest eccentricity encounter differing from the others.  By simulating a modest but non-zero velocity at infinity, we perhaps underestimate the capture rates for the low velocity tail of the distribution, and overestimate the high velocity capture rate.  Since the dominant contribution to the capture rate comes from relative velocities $\leq 2$ \kmss for all masses, the net contribution of this simplification is not thought to significantly alter the capture rate estimations. 

\begin{figure}
 \centering
  \plotone{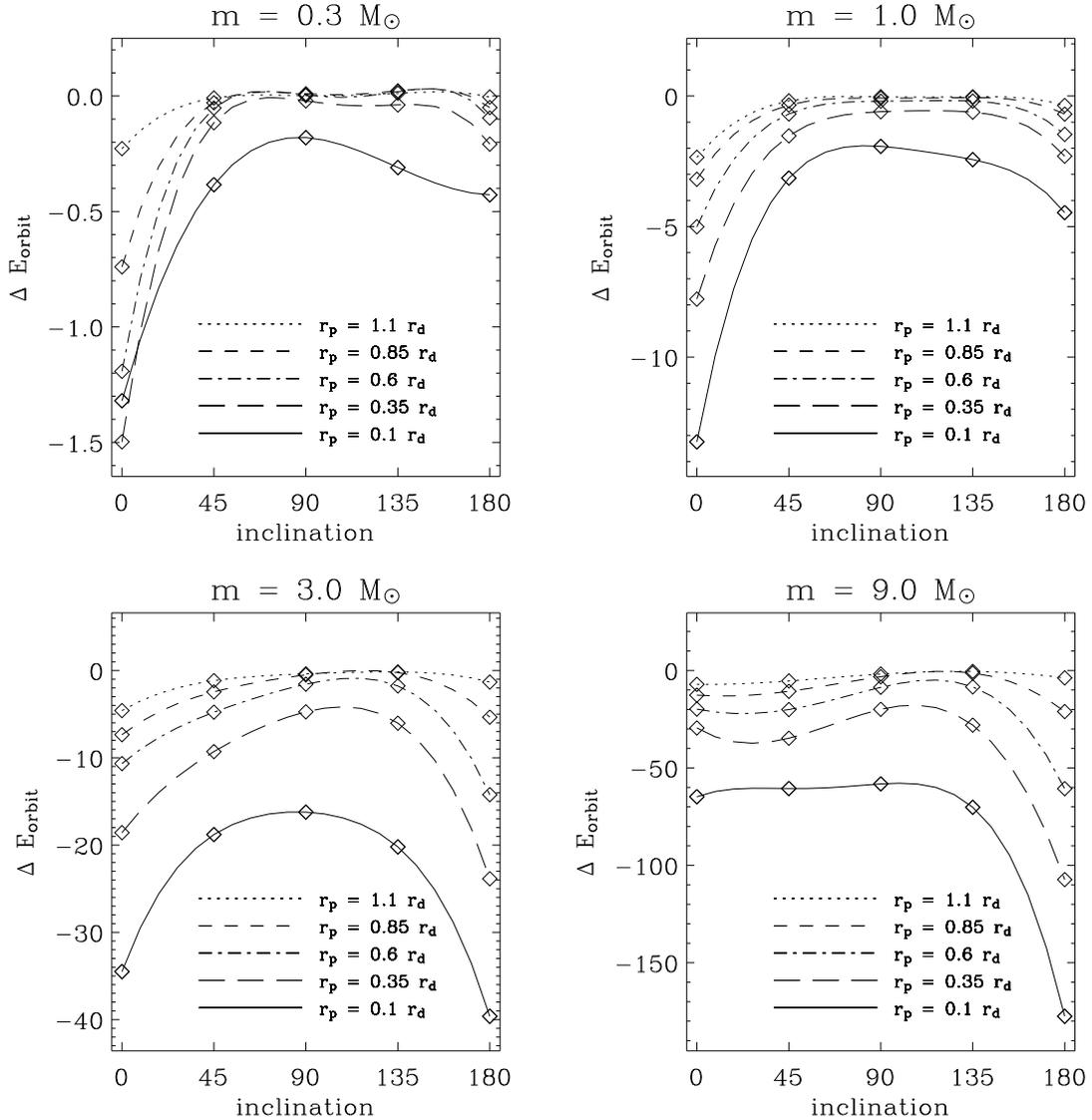}
  \caption{Change in energy as a function of inclination for each periastron radius $r_{p}$ and impactor mass $m$.  The diamonds are the simulation data points, and the curves are the interpolated values used in calculating binary formation rates.  The energy is in code units, where mass is in \msun, velocity is in \kms, and distance is in AU.}
  \label{binary_vals}
\end{figure}

We consider four different masses $m$ for the impactor: 0.3, 1.0, 3.0, and 9.0 \msun.  We cover the range of inclination angles [0,180] in 45 degree increments, and we simulate encounters with unperturbed periastra $r_{p}$ of 0.1, 0.35, 0.6, 0.85, and 1.1 $r_{d}$.  This provides a grid of 100 encounters.  At the end of the simulation we sort the gas particles into three groups: those bound energetically to the impactor, those ejected from the system, and those bound to the primary.  The stars and their associated gas particles are treated as a point mass at the center of mass of the grouping, and these masses are used to determine an orbital energy.  A representative encounter is shown in figure \ref{encounter_example}, with $m = 3.0$ \msun, $r_{p} = 0.35 r_{d}$, and inclination angle $i = 45$ degrees.  In the right panel, the impactor has finished its passage through the disk.  We run each encounter until the orbital energy has stopped evolving.  The results of these simulations are summarized in figure \ref{binary_vals}.  

The change in energy for a given inclination generally decreases with increasing periastron distance, a result in agreement with \citet{hel95} and \citet{bof98}.  There are also differences between the low mass and high mass impactors, best seen by comparing the change in energy associated with prograde versus retrograde collisions for $r_{p} = 0.1r_{d}$.  Note that the disks considered here are gaseous; several authors have considered disk interactions with purely gravitational effects included.  \citet{pfa03} showed that for low mass disks and non-penetrating encounters, ignoring hydrodynamical forces is a valid simplification, and uses this to explore a large parameter space in \citet{pfa04}.  \citet{hal96} used a reduced three-body approach to explore encounters in the absence of hydrodynamics or disk self gravitation, to identify the regions of the disk that dominate energy transfer.  

However, with more massive disks and penetrating encounters, orbital energy loss in retrograde encounters is dominated by the dissipation of energy in the trailing shock behind the intruding star \citep{hel95,bof98,moe06}.  As the mass of the impactor increases, this mechanism becomes more effective.  For $m = 0.3$ \msun, the prograde encounter is over three times more dissipative than the retrograde; for $m = 9.0$ \msun, the retrograde case is over twice as dissipative as a prograde passage.  As the periastron radius increases, this mechanism plays less of a role, until at $r_{p} = 1.1 r_{d}$ it is less effective than the spiral arms set up in a prograde encounter.  The capture of a mass by the impactor plays a small role in the change of orbital energy.  The lower mass impactors, $m \leq 1.0$ \msun, capture very little mass from the disk.  The more massive impactors can increase their mass by as much as 10 percent during a passage, but the change in orbital energy is typically an order of magnitude larger than can be accounted for by simply adding mass to the impactor in a momentum conserving fashion.

\section{DETERMINING CAPTURE RATES}
In order to calculate the efficiency of disk-capture as a binary formation mechanism, we largely follow the method of \citet{hel95}, also used by \citet{bof98}.  We modify the method to account for the difference in mass between the impactor and the primary, and allow for a mass dependent velocity dispersion.  We describe the method here.  

Determining binary capture rates is closely related to the calculation of stellar collision rates found in \citet{bin87}.  Instead of determining the frequency of encounters with closest approach less than a stellar radius, we are interested in the frequency of passages that have a closest approach within some critical radius $r_{c}$, within which the energy dissipated via disk interactions is sufficient to form a binary.  This radius depends on the stellar masses, initial orbital energy, and inclination angle of the encounter.  For a cluster of equal mass stars in a Maxwellian velocity distribution, the rate of encounters with a closest approach radius less than $r$ is given by
\begin{equation}
  \label{bin_trem_rate}
  \gamma_{BT} = 4 \sqrt{\pi} n \sigma r^{2} \left(1 + \frac{G m_{\star}}{\sigma^{2} r} \right),
\end{equation}
with $n$ the number density of the cluster, $m_{\star}$ the mass of the stars and $\sigma$ the one dimensional velocity dispersion \citep{bin87}.  

The cross section for such an encounter is easily generalized to the case of uneven mass stars $M$ and $m$, with velocities velocities ${\bf V}$ and ${\bf v}$,
\begin{equation}
  \label{cross_section}
  \Sigma(r,\mathcal{M},v_{rel}) = \pi r^{2} + \frac{2 \pi G \mathcal{M} r}{v_{rel}^{2}},
\end{equation}
with $\mathcal{M} = M + m$ the total mass of the system, and $v_{rel} = |{\bf V} - {\bf v}|$.  For calculating capture rates, the relevant encounter radius is not fixed, but depends on $m$, $v_{rel}$, and inclination angle $i$.  Via interpolation of the data in figure \ref{binary_vals} we can find a critical radius $r_{c}$, such that the encounter satisfies
\begin{equation}
  \Delta E_{orbit} < - \frac{1}{2} \mu v_{rel}^{2},
\end{equation}
with $\mu$ the reduced mass of the system, $Mm/\mathcal{M}$.
The cross section for capture is then defined as
\begin{equation}
  \label{capture_cross_section}
  \Psi(i,\mathcal{M},v_{rel}) = \Sigma(r_{c},\mathcal{M},v_{rel}).
\end{equation}

 The capture rate for a star with mass $M$ and velocity ${\bf V}$ is then obtained by multiplying the cross section by $v_{rel}$ and integrating over the velocity distribution of the encounter partners $f({\bf v})$; an average over the primary velocity ${\bf V}$ and inclination angle $i$ is obtained by 
\begin{equation}
  \label{gamma_raw}
  \Gamma_{c}(m) = \frac{\int_{0}^{\infty}\int_{0}^{\infty}\int_{0}^{\pi}f({\bf V})f({\bf v})\Psi(v_{rel},i,\mathcal{M})v_{rel}sin(i)di~d^{3}{\bf v}~d^3{\bf V}}{\int_{0}^{\infty}\int_{0}^{\pi}f({\bf V})sin(i)di~d^{3}{\bf V}}.
\end{equation}

We assume a Maxwellian velocity distribution with a one dimensional, mass dependent velocity dispersion of the form
\begin{equation}
\label{dispersion}
  \sigma^{2} = \left( \frac{m_{0}}{m} \right)^{\alpha}\sigma_{0}^{2},
\end{equation}
where $m_{0}$ and $\sigma_{0}$ are chosen to match some observations, for instance $m_{0} =0.3$ \msun, and $\sigma_{0}\sim2.3$ \kmss would be appropriate for the Orion Trapezium cluster \citep{alt88,jon88}.

After applying this velocity distribution to equation \ref{gamma_raw}, we introduce new integration variables ${\bf \tilde{V}} = {\bf V} - {\bf v}$, and ${\bf \tilde{v}} = (M^{\alpha}{\bf V} + m^{\alpha}{\bf v})/(M^{\alpha}
+ m^{\alpha})$.  With this change we have
\begin{eqnarray}
  \label{gamma_nointegral}
  \Gamma_{c}(m) = \frac{n (Mm)^{3\alpha/2}}{16 \pi^{3} m_{0}^{3\alpha} \sigma_{0}^{6}} \int_{0}^{\infty} \int_{0}^{\infty} \int_{0}^{\pi}&&exp\left[-\frac{(M^{\alpha}+m^{\alpha})^{2}\tilde{v}^{2} + (Mm)^{\alpha}\tilde{V}^{2}}{2 m_{0}^{\alpha}\sigma_{0}^{2}(M^{\alpha}+m^{\alpha})}\right] \times \\ 
  &&\Psi(\tilde{V},i,\mathcal{M}) \tilde{V}sin(i)di~d^{3}{\bf \tilde{v}}d^{3}{\bf \tilde{V}}.
\end{eqnarray}

The integral over ${\bf \tilde{v}}$ is straight forward, and we arrive at a formula for the capture rate of a star with mass $M$ interacting with stars of mass $m$,
\begin{eqnarray}
\label{gamma}
  \Gamma_{c}(m) = \frac{n (Mm)^{3\alpha/2}}{\sqrt{2\pi} m_{0}^{3\alpha/2} \sigma_{0}^{3}(M^{\alpha}+m^{\alpha})^{3\alpha/2}} \int_{0}^{\infty} \int_{0}^{\pi}&&exp\left[-\frac{(Mm)^{\alpha}\tilde{V}^{2}}{2 m_{0}^{\alpha}\sigma_{0}^{2}(M^{\alpha}+m^{\alpha})}\right] \times\\ &&\Psi(\tilde{V},i,\mathcal{M})\tilde{V}^{3}sin(i)di~d\tilde{V}.
\end{eqnarray}

For the case of no mass dependence in the velocity dispersion, $\alpha = 0$, this reduces to
\begin{equation}
\label{gamma_alpha_zero}
  \Gamma_{c}(m) = \frac{n}{4\sqrt{\pi}\sigma_{0}^{3}} \int_{0}^{\infty} \int_{0}^{\pi}exp\left(-\frac{\tilde{V}^{2}}{4\sigma_{0}^{2}}\right) \Psi(\tilde{V},i,\mathcal{M})\tilde{V}^{3}sin(i)di~d\tilde{V},
\end{equation}
which is equivalent to the formula used by \citet{bof98}.

If instead of using the experimentally found $\Psi(i,\mathcal{M},v_{rel})$ we use the analytical $\Sigma(r_{c},\mathcal{M},v_{rel})$ in equation \ref{gamma_raw}, the inclination dependence disappears and the integral over ${\bf \tilde{V}}$ is easily done, resulting in a generalized collision rate, 
\begin{equation}
\label{our_rate}
  \gamma = 2 \sqrt{\frac{2 \pi (M^{\alpha}+m^{\alpha})}{(Mm)^{\alpha} }}  n r^{2} m_{0}^{\alpha/2} \sigma_{0}  \left[ 1 + \frac{G \mathcal{M}}{\sigma_{0}^{2} r}
  \left(\frac{M}{m_{0}}\right)^{\alpha}
  \frac{m^{\alpha}}{M^{\alpha}+m^{\alpha}} \right].
\end{equation}

For the special cases $\alpha = 1$ (energy equipartition) and $\alpha = 0$ (no mass dependence in the velocity dispersion), equation \ref{our_rate} reduces to
\begin{equation}
\label{our_rate_one}
  \gamma_{\alpha=1} = 2 \sqrt{\frac{2 \pi m_{0}}{\mu}}n \sigma_{0} r^{2} \left(1 + \frac{G \mathcal{M} \mu}{ m_{0} \sigma_{0}^{2} r} \right),
\end{equation}

\begin{equation}
\label{our_rate_zero}
  \gamma_{\alpha=0} = 4 \sqrt{\pi}n \sigma_{0} r^{2} \left(1 + \frac{G \mathcal{M}}{ 2\sigma_{0}^{2} r} \right),
\end{equation}
Equation \ref{our_rate_zero} is equivalent to equation \ref{bin_trem_rate}, with the mass $m_{\star}$ replaced by the average of $M$ and $m$.

Equation \ref{gamma_nointegral} is the capture rate for a single star of mass $M$ awash in a sea of stars with uniform mass $m$.  To arrive at a total capture rate for a specific star with mass $M$ in a cluster with density $n$ and some mass function $\xi(m)$, we average equation \ref{gamma_nointegral} over the mass range $m_{l}$ to $m_{u}$, 
\begin{equation}
\label{avg_rate}
  \Gamma_{c} = \frac{\int_{m_{l}}^{m_{u}} \xi(m)\Gamma_{c}(m) dm}{\int_{m_{l}}^{m_{u}}\xi(m) dm}.
\end{equation}

In this work we use the IMF from \citet{kro01},
\begin{equation}
\label{imf}
  \xi(m) = \left\{
  \begin{array}{cc} 
    C_{1}~m^{-0.3+\beta}, & 0.01 \le m/M_{\sun} < 0.08, \\ 
    C_{2}~m^{-1.3+\beta}, & 0.08 \le m/M_{\sun} < 0.5, \\
    C_{3}~m^{-2.35+\beta}, & 0.5 \le m/M_{\sun} < M_{max}, \\
  \end{array}
  \right.
\end{equation}
where we have added a constant $\beta$ to all the exponents; this constant is included to approximate the effects of mass segregation in a cluster.  The constants $C_{i}$ enforce continuity at the transitions between the different power laws, and are functions of $\beta$.  If $\beta = 0$, we have the unmodified, present day IMF.  As $\beta$ increases, skewing the mass function toward higher values, the average stellar mass increases and encounter rates are altered.  The effect of $\beta$ on the mass function is summarized in table \ref{mass_segregation_table}.  Note that this single parameter alteration of the mass function is insufficient to reproduce the mass function generated by dynamical mass segregation \citep{gur04}.  We use this to illustrate the character of mass segregation effects in a simple fashion.  The maximum mass $M_{max}$ depends on the cluster mass, which sets the maximum stellar mass in the cluster \citep{wei06}.  While we consider the primary to be the most massive star in the cluster, because we only simulated encounters with companions up to 9 \msuns we set $M_{max} = 9$ \msun.  We now have the binary formation rate averaged over the IMF, and as a function of the level of mass segregation and mass dependent velocity dispersion.
\begin{deluxetable}{ccccc} 
\tablecolumns{5} 
\tablewidth{0pc} 
\tablecaption{The effect of the constant $\beta$ on the IMF.\label{mass_segregation_table}} 
\tablehead{ 
\multicolumn{1}{c}{} & \multicolumn{4}{c}{$\beta$}  \\
\cline{2-5}  
\colhead{Mass Range}    &  \colhead{0.0} &   \colhead{0.5}   & 
\colhead{0.75}  &  \colhead{1.0} \\ 
\multicolumn{1}{c}{(\msun)} & \multicolumn{4}{c}{(\% of stars)} 
}
\startdata 
0.3-0.5 & 42.9 & 30.1 & 23.5 & 17.2 \\ 
0.5-1.0 & 35.0 & 32.5 & 29.2 & 24.6 \\ 
1.0-9.0 & 21.4 & 34.3 & 41.5 & 48.1 \\ 
9.0-20.0 & 0.77 & 3.08 & 5.77 & 10.1 \\ 
\cline{1-5}
$\bar{m}/$\msun & 0.32 & 0.88 & 1.66 & 3.37
\enddata
\tablecomments{See equation \ref{imf} for the IMF and definition of $\beta$.  The average masses $\bar{m} = \int_{m_{l}}^{m_{u}} m \xi(m) dm / \int_{m_{l}}^{m_{u}}\xi(m) dm$ are integrated over the range 0.3 - 20.0 \msun.}
\end{deluxetable}

\begin{figure}
 \centering
  \plotone{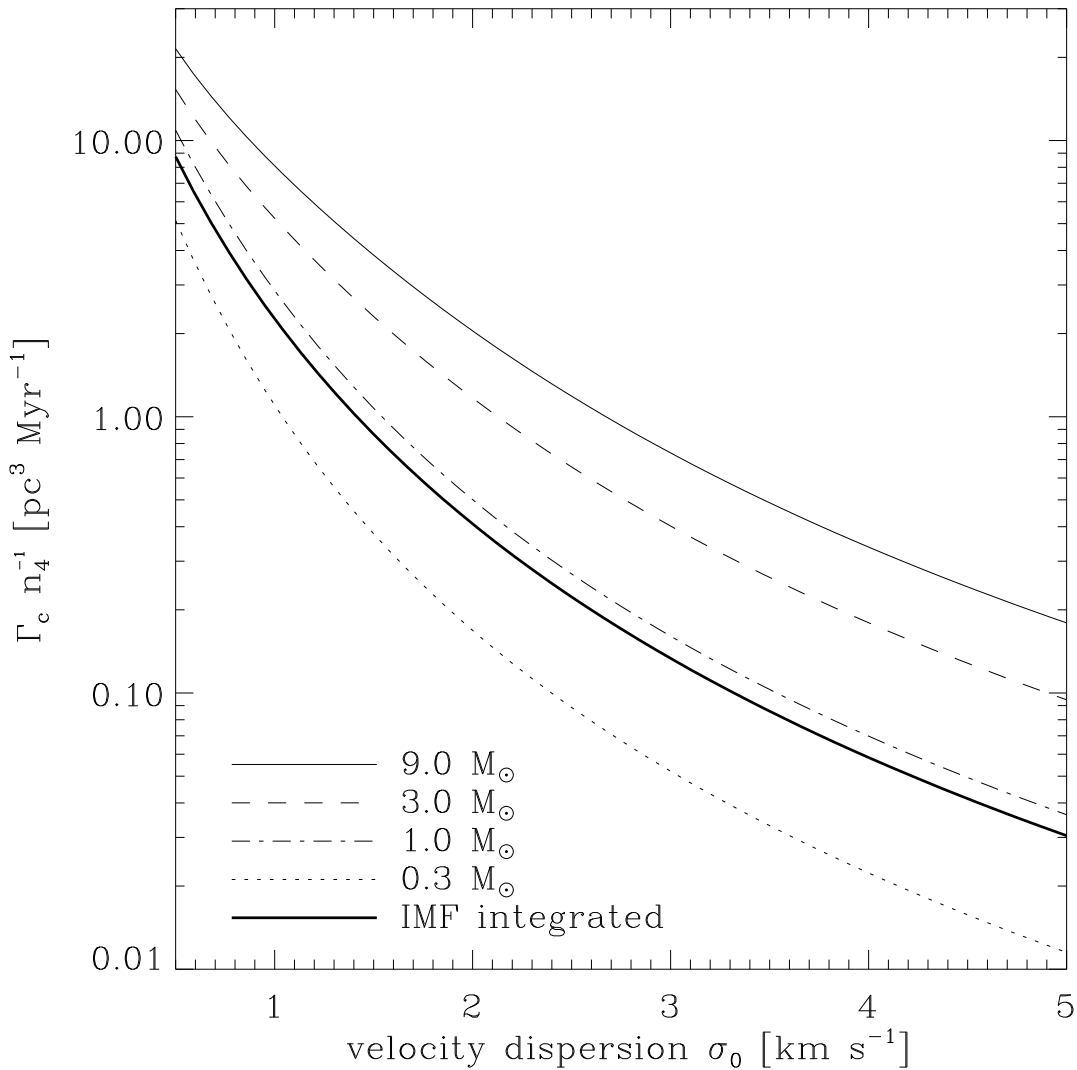}
  \caption{Binary formation rates normalized to a cluster with number density $n_{4} = 10^{4}$ pc$^{-3}$.  Shown are the rates for a 20 \msuns star with a 2 \msun, 500 AU disk in a cluster of stars with uniform mass 0.3, 1.0, 3.0, and 9.0 \msun, and in a cluster the IMF of equation \ref{imf} for the range $0.3 \leqq m/M_{\sun} \leqq 9.0$.  In this plot there is no mass dependence in the velocity dispersion, nor mass segregation in the mass function.}
  \label{binary_rates_log}
\end{figure}

Figure \ref{binary_rates_log} shows the capture rates, with no mass dependence in the velocity dispersion, for the cases where all the cluster members (besides the capturing star) are 0.3, 1.0, 3.0, and 9.0 \msun, as well as the rate averaged over the IMF.  Because of the range of masses we simulated, we take $m_{l} = 0.3$ \msuns and $m_{u} = 9.0$ \msun.  At the high end of the mass spectrum this truncated mass function is fairly complete for $\beta \lesssim 0.5$, as stars with masses $m > 9.0$ \msuns contribute $\lesssim 3$ percent of the total population.  For higher values of $\beta$ we are missing a significant portion of the mass function, which leads to an underestimation of the capture rates.  The rates are normalized to a cluster with stellar density $n_{4} = 10^{4}$ pc$^{-3}$, and can be compared to figure 8 in \citet{bof98} and figure 5 in \citet{hel95}.  The IMF integrated rate calculated here is roughly an order of magnitude higher than those previously calculated for encounters between lower mass stars, with mass ratios $M/m \geqq 0.5$.

By ignoring any encounters which take place outside the range $0.1~r_{d} \leq r_{p} \leq 1.1~r_{d}$ we underestimate the rates somewhat; \citet{bof98} find that some encounters out to $r_{p} = 2.0~r_{d}$ result in a binary.  These weak encounters are most important at low relative velocities.  Since we are most interested in scenarios with $\sigma_{0} \sim 2.0$ \kms, the exclusion of encounters with periastra greater than those simulated is negligible.  The frequency of encounters with $r_{p} < 0.1~r_{d}$ is small enough that ignoring them is a small effect.  \citet{hel95} finds that extrapolating to direct collisions from $r_{p} = 0.2~r_{d}$ results in changes to the capture rate on the order of 10 percent, and \citet{bof98} reports similar results extrapolating inward from $r_{p} = 0.5~r_{d}$.

\section{DISCUSSION OF RESULTS}
\subsection{Comparison to Previous Work}
In this section we consider the case of no mass dependence in the velocity dispersion, and no mass segregation in the IMF, i.e. $\alpha = \beta = 0$.
Table \ref{comparison_table} shows the capture rates from this work, for a variety of star forming environments, compared to rates calculated in previous work.  For the Trapezium center we take $n = 2 \times 10^{4}$ pc$^{-3}$ \citep{hil98}, and $\sigma_{0} = 2.3$ \kmss \citep{alt88,jon88}.  The other regions considered are those in \citet{bof98}.  For a massive star in a cluster of lower mass companions, the capture rates are roughly an order of magnitude higher than encounters involving impactors with masses $M/2 \leq m \leq 2M$.  

\begin{deluxetable}{lccccc} 
\tablecolumns{6} 
\tablewidth{0pc} 
\tablecaption{Capture rates in different environments.\label{comparison_table}} 
\tablehead{ 
\multicolumn{3}{c}{} & \multicolumn{3}{c}{References} \\
\cline{4-6}
\multicolumn{3}{c}{} & \colhead{This Work}  &  \colhead{1}  &  \colhead{2}\\
\colhead{Region}    &  \colhead{$n$} &   \colhead{$\sigma$}   & 
\multicolumn{3}{c}{$\Gamma_{c}$}\\
\colhead{}    &  \colhead{(pc$^{-3}$)} &   \colhead{(\kms)}   & 
\multicolumn{3}{c}{(Myr)}
}
\startdata 
Trapezium, center & $2 \times 10^{4}$ & 2.3 & 0.53 & 0.06 & 0.018 \\ 
Trapezium & $2 \times 10^{3}$ & 1.5 & 0.17 & 0.02 & $5 \times 10^{-3}$ \\ 
Open cluster & $10^{2}$ & 1 & 0.02 & $3 \times 10^{-3}$ & $8 \times 10^{-4}$  
\enddata 
\tablerefs{(1) Boffin et al. 1998; (2) Heller 1995.}
\end{deluxetable} 

\begin{figure}
 \centering
  \plotone{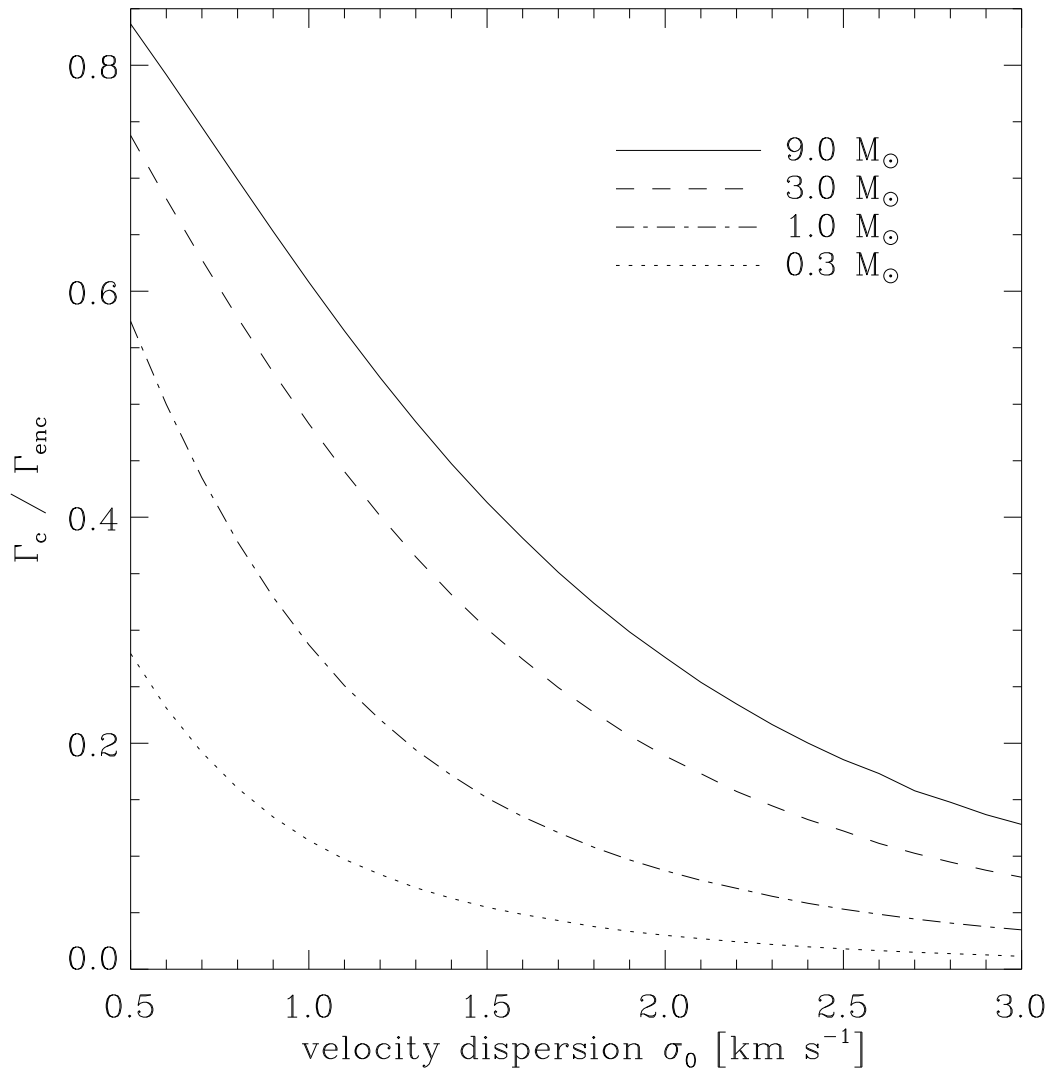}
  \caption{Probability of a random encounter resulting in a binary, for each of the impactor masses simulated.  $\Gamma_{c}$ is the binary formation rate, and $\Gamma_{enc}$ is the rate of all encounters within $1.1~r_{d}$.}
  \label{binary_probabilities}
\end{figure}

Figure \ref{binary_probabilities} shows the probability of an average encounter resulting in a captured binary for each of the impactor masses.  The probability is given by $\Gamma_{c}(m)/\Gamma_{enc}$, with $\Gamma_{enc}$ the total encounter rate for $r < 1.1~r_{d}$, found by averaging equation \ref{our_rate_zero} over the IMF.  The probabilities are one to two orders of magnitude larger than those of \citet{hel95} or \citet{bof98}.  However, despite the lowest mass impactors, $m_{i} = 0.3$ \msun, encountering a disk nearly an order of magnitude more massive than themselves, they have the lowest probability of capture.  The increased gravitational radius of the more massive impactors is enough to overwhelm any effect such a relatively massive disk might have on a low mass impactor.

The increased effectiveness of disk interactions in capturing companions seen here is due to two different effects.  Due to their greater gravitational focusing, massive stars in a cluster of lower mass companions will have a larger encounter rate, which though not the main reason for the larger rates here, contributes somewhat.  The main cause of these higher rates lies in the ratio of the disk mass to its radius.  The primary-disk systems in this work have the same ratio $m_{d}/m_{p} = 0.1$ as \citet{hel95}.  However, while these disks are 20 times as massive as Heller's, the radii are only 4 times as large.  The disks in Boffin et al. are both less massive and have larger radii. It is this higher relative concentration of disk material that is the dominant cause of the high rates calculated here.

In lower density regions such as an open cluster, the increased efficiency of high mass capture is insufficient to make it a significant contributer to the total binary fraction.  However, in dense regions such as the Trapezium cluster in Orion, with stellar density $\sim 2 \times 10^{4}$, one could expect more than 50\% of the massive stars to have captured a companion after 1 Myr.  This is a high enough rate to be considered a real contribution to the observed multiplicity of high mass stars.  

In the Arches cluster near the Galactic center, the central mass density is $\sim3\times10^{5}$ \msuns pc$^{-3}$ \citep{fig99}, compared to the Trapezium core at $\sim3\times10^{4}$ \msuns pc$^{-3}$ \citep{hil98}.  At such high densities the capture rate $\Gamma_{c}$ could be well above unity after 1 Myr, though the only limit on the velocity dispersion is $\sigma < 22$ \kmss \citep{fig02}, which would drive the rate down to negligible values.  It is worth noting that the majority of encounters that result in capture occur at low relative velocities $v_{rel} \lesssim 4$ \kms.  As the velocity dispersion increases and such encounters are less frequent, the capture rate plummets, clearly seen in figure \ref{binary_rates_log}.

\subsection{Effect of a Mass Dependent Velocity Dispersion}
\begin{figure}
 \centering
  \plottwo{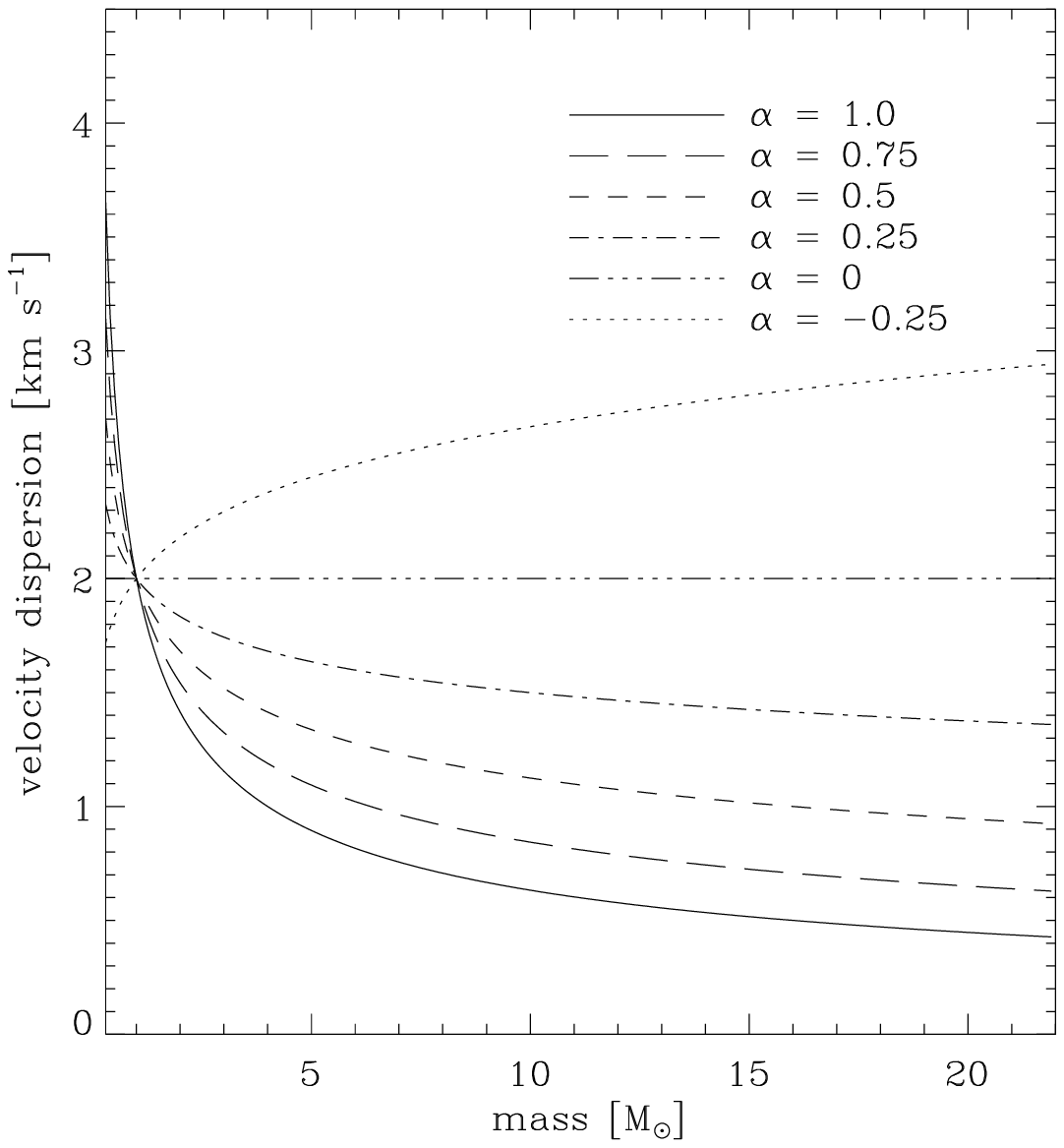}{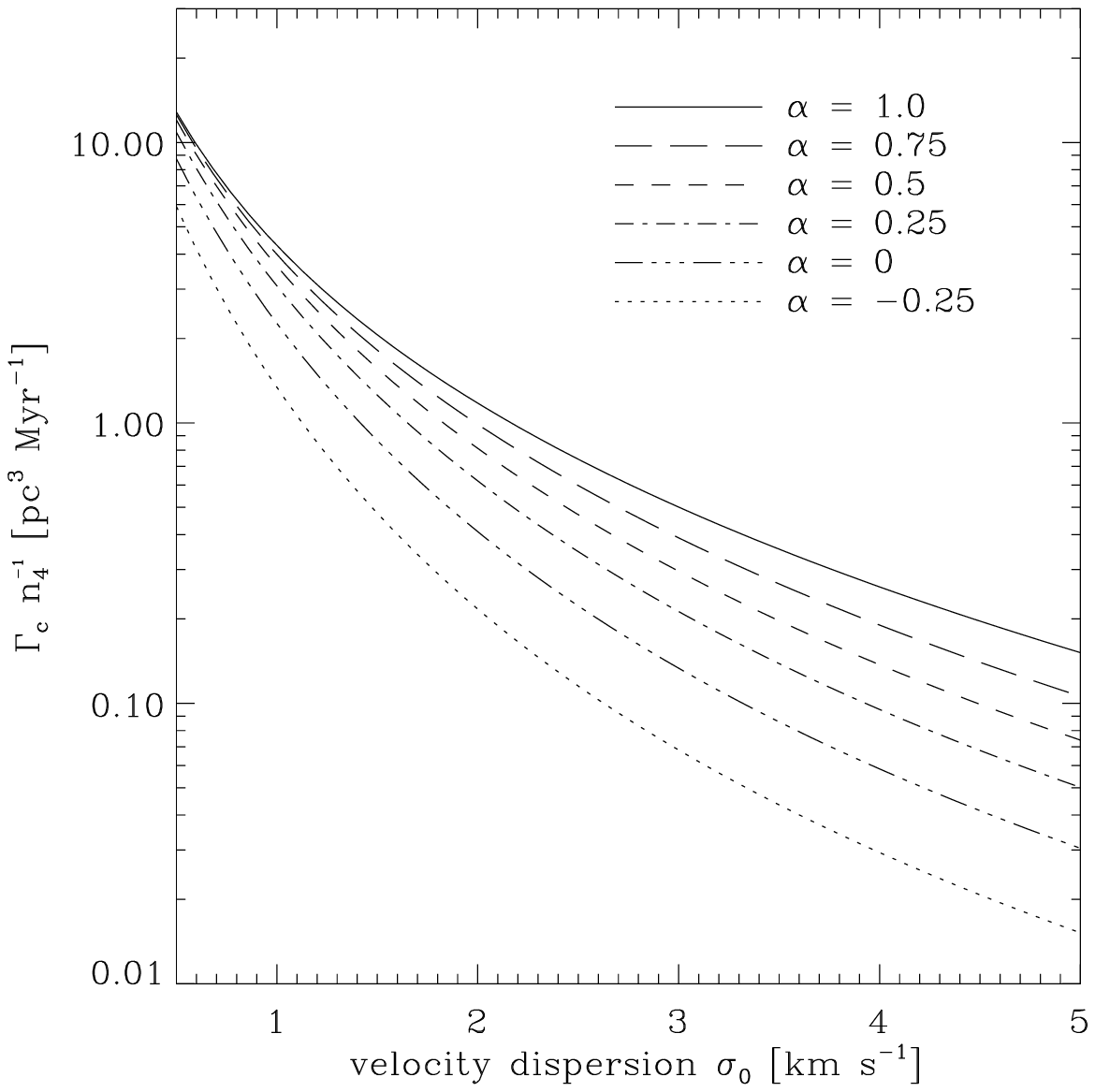}
  \caption{{\it Left}: The velocity dispersion (equation \ref{dispersion}) as a function of mass for different values of $\alpha$, the mass-dependence parameter.  The dispersion is anchored at $m_{0} = 1.0$ \msun, $\sigma_{0} = 2.0$ \kms.  {\it Right}: The IMF integrated capture rate for the same $\alpha$ values.  The dispersion $\sigma_{0}$ is allowed to vary, with $m_{0}$ again 1.0 \msun.}
  \label{binary_rates_alpha}
\end{figure}

Shown in figure \ref{binary_rates_alpha} are the IMF integrated capture rates, for different values of $\alpha$ in equation \ref{dispersion}, as well as the velocity dispersions as a function of mass.  The results below depend not only on the value of $\alpha$ but on $m_{0}$, the anchor point of the velocity dispersion.  We choose $m_{0} = 1.0$ \msuns so that  the dispersion on the low end, $m_{\star} \sim 0.3$ \msun, is not unrealistically high, yet a good differentiation between the most massive and least massive stars' dispersions can be seen.

As the cluster tends toward equipartition, $\alpha = 1$, the binary formation rate roughly triples.  The cause of this is that the massive stars are moving more slowly, and the lower relative velocity increases the encounter rate and the probability of capture.  For the case of $\alpha = -0.25$, with the more massive stars moving faster, the rate drops by roughly a factor of two.

A cluster with stars in energy equipartition will tend to have more captured binaries involving its most massive stars.  However, in most clusters equipartition is by no means a good assumption.  Observationally, there is evidence in Orion's Trapezium cluster for a weak trend toward increasing velocity dispersion with decreasing mass \citep{jon88,hil98}.  Hillenbrand \& Hartmann find $\sigma=2.81$ \kmss for $0.1 < M/M_{\sun} < 0.3$, and $\sigma=2.24$ \kmss for $1.0 < M/M_{\sun} < 3.0$.  In our formula for the mass dependence of the velocity dispersion, a value of $\alpha \sim 0.2$ seems appropriate for the Trapezium.  At such low values, the increase in the capture rate is only $\sim 25$ percent, much less dramatic than a cluster in equipartition.      

\subsection{Effect of Mass Segregation}
An effect which would raise the rates calculated here is mass segregation; there is evidence for this in Arches and Trapezium clusters.  In the Arches cluster, \citet{sto02} find that the overall slope of the IMF flattens inside 5\arcsec, from a value of $-2.7 \pm 0.7$ to $-1.1 \pm 0.3$.  In Trapezium, the mass function of the cluster taken as a whole ($r < 2.5$ pc) is similar to a standard IMF \citep{hil97}.   \citet{hil98} find strong evidence for mass segregation toward the cluster core for masses down to $\sim 1$ \msun, and find that within 0.05 pc of the cluster core the mean mass per star $\bar{m} \sim 2-4$ \msun, compared to $\bar{m} \sim .75$ \msuns for $r > 0.1$ pc.  We approximate the effects of segregation by increasing the value of $\beta$ in equation \ref{imf}.  In order to raise the average stellar mass above 1 \msun, we need $\beta \sim 0.75$, though for the Arches cluster $\beta = 1$ would be more appropriate.

\begin{figure}
 \centering
  \plottwo{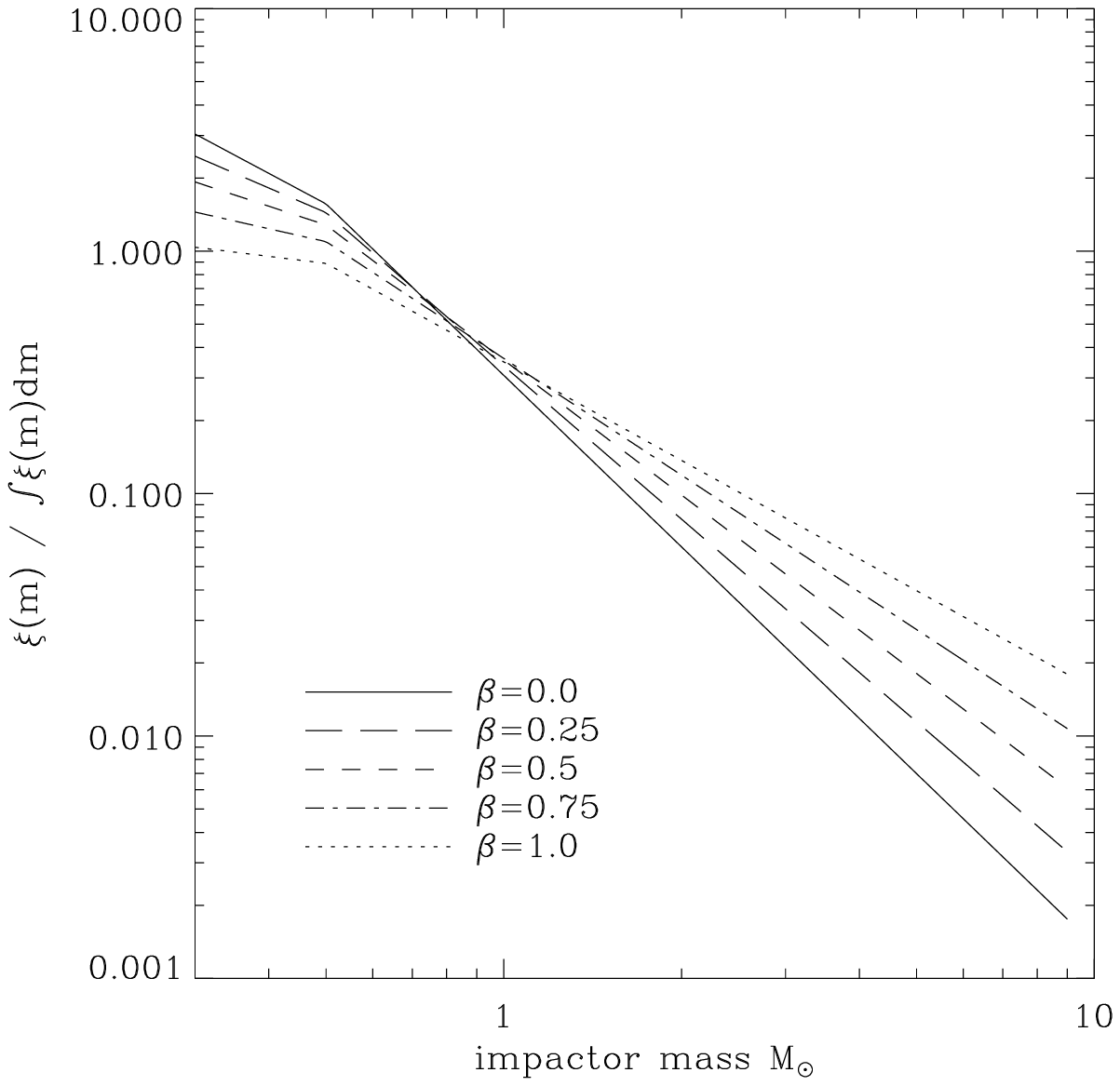}{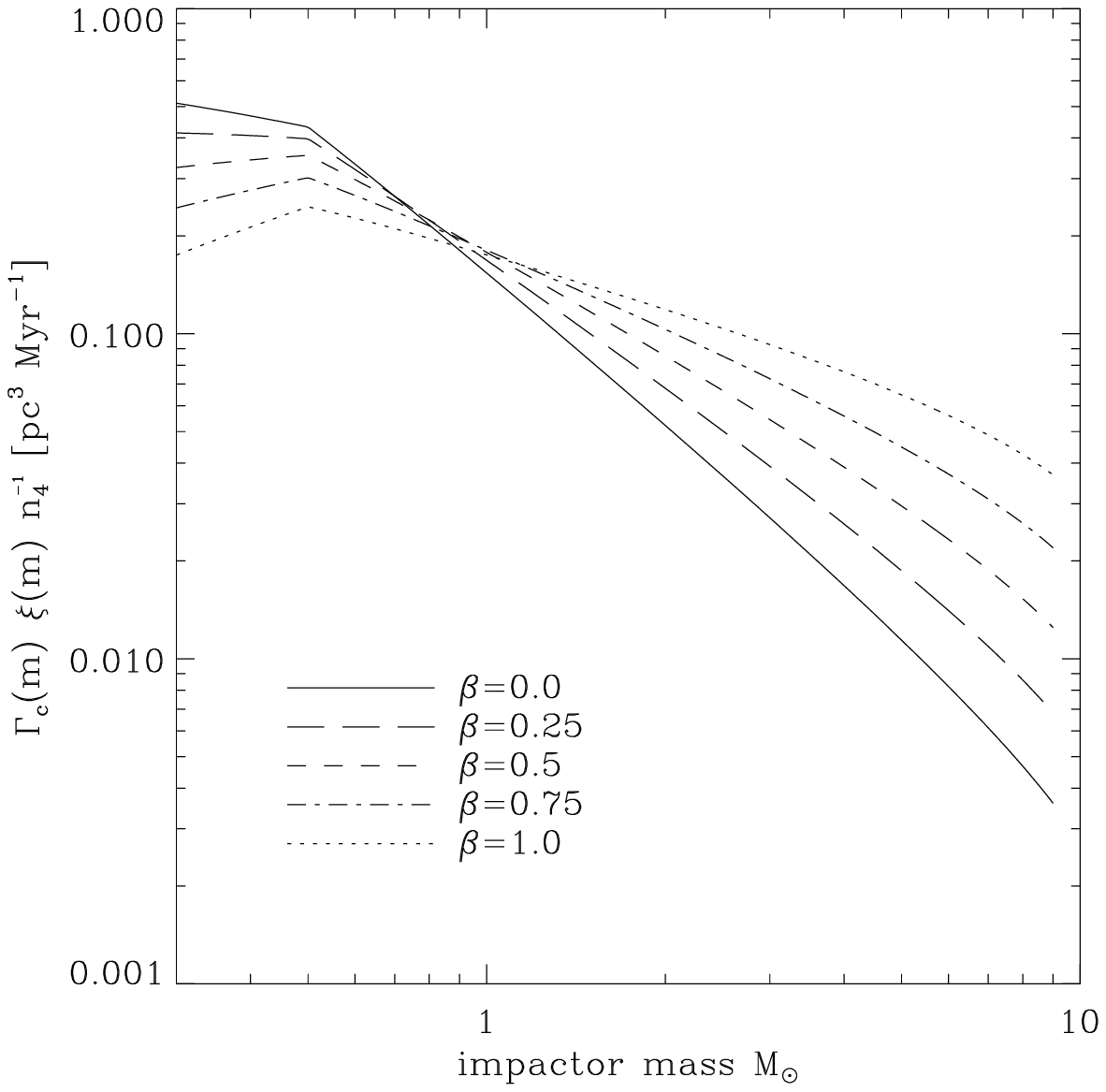}
  \caption{{\it Left}: The mass function of equation \ref{imf} for different values of $\beta$, the mass-segregation parameter.  {\it Right}: The capture rate as a function of impactor mass for different values of $\beta$.  The velocity dispersion is set at 2 \kms.}
  \label{binary_rates_beta_log}
\end{figure}

As $\beta$ increases from 0, the more massive stars become over-represented in the mass function compared to the present-day IMF.  In the absence of a mass dependent velocity dispersion, the total binary formation rate essentially follows the rate that would be found if all the stars had the average stellar mass.  What is more interesting is the capture rate broken down by mass.  Shown in figure \ref{binary_rates_beta_log} are the mass functions for different values of $\beta$, and the capture rates as a function of mass weighted by the mass function.  The velocity dispersion $\sigma_{0} = 2$ \kms.    

As $\beta$ approaches increases from 0 to 1, the mass function flattens and 9 \msuns stars increase their representation in the stellar population by over an order of magnitude.  For $\beta = 0$ the ratio $\xi(0.5)/\xi(9.0) \sim 850$.  For $\beta = 1$, $\xi(0.5)/\xi(9.0) \sim 50$.  This change in the proportion of massive stars is reflected directly in the capture rates.  Since massive stars have higher capture probabilities (see figure \ref{binary_probabilities}) the ratios of capture rates are smaller.  For $\beta = 0$ the rate for 0.5 \msuns impactors is $\sim 50$ times as large as the rate for 9.0 \msuns stars; at $\beta = 1$ the rate for 0.5 \msuns stars is only $\sim 7$ times as high as that of 9.0 \msuns stars.  If mass segregation occurs in a young cluster on a timescale short enough that disk capture has a chance to contribute to the multiplicity of massive stars, presumably found near the center of the cluster, than capturing relatively massive companions would be likely.  In an environment like the Arches cluster, where the mass function is in fact flat, a massive companion is the most likely outcome of a capture event.  Mass segregation thus has the double effect of increasing the net capture rate, and biasing the mass of captured stars toward the high end of the mass spectrum.

\subsection{The Fate of the Binaries}
\label{fate}
With semi-major axes of several $\times 10^{3}$ AU, the resultant binaries from these captures present extremely large cross sections for further interactions with other cluster members, especially in the dense ($n \sim 10^{3} - 10^{4}$ pc$^{-3}$) environment of a young, rich star cluster.  If binaries formed via disk encounters are to contribute to the observed high multiplicity of massive stars, they need to be robust enough to survive.

A full treatment of this question would include a campaign of n-body simulations, beyond the scope of this paper.  In order to get some sense of the fate of captured binaries, we instead compare the energy of an encounter with another cluster member to the orbital energy of the binaries.  In considering an encounter between a binary and a single star there is a critical relative velocity, above which the system's total energy is positive, and total ionization is possible \citep[see e.g.][]{heg75}.  For a single star of mass $m_{2}$ encountering a binary with total mass $\mathcal{M}$, reduced mass $\mu$ and semi-major axis $a$, this relevant comparison of energies is 
\begin{equation}
  \label{critical_velocity}
    \frac{1}{2} \frac{\mathcal{M} m_{2}}{\mathcal{M} + m_{2}} v_{c}^{2} = \frac{G\mathcal{M}\mu}{2 a}.
\end{equation}
The term on the left is the energy associated with the encounter between the field star and the binary, moving with relative velocity $v_{c}$.  The term on the right is the negative of the orbital energy of the binary.  For relative velocities greater than $v_{c}$, ionization is energetically permitted.  

For velocities below $v_{c}$, these encounters can be resonant, with complex orbits eventually leading the ejection of one of the stars, or a relatively simple exchange wherein a new binary is formed \citep{hut83,heg03,fre04}.  The critical velocity $v_{c}$ is not a hard boundary, but for velocities above $\sim 2 v_{c}$, ionization is the dominant result.  In the case of an encounter which forms a temporary, unstable three body system, the least massive member is preferentially ejected \citep{ano86,mik86,bat02}.  Note, though, that this ``dynamical biasing" is not a rule, especially with soft binaries \citep{heg75}.   Additionally, most numerical studies of binary-single star interactions focus on nearly equal mass components, and it is not immediately clear that the results apply to a situation with one dominant mass.  However, as a rough guide we continue the analysis, with the assumption that some (if not most) of the time, an encounter with another cluster member will leave the most massive star in a binary.   

\begin{figure}
 \centering
  \plotone{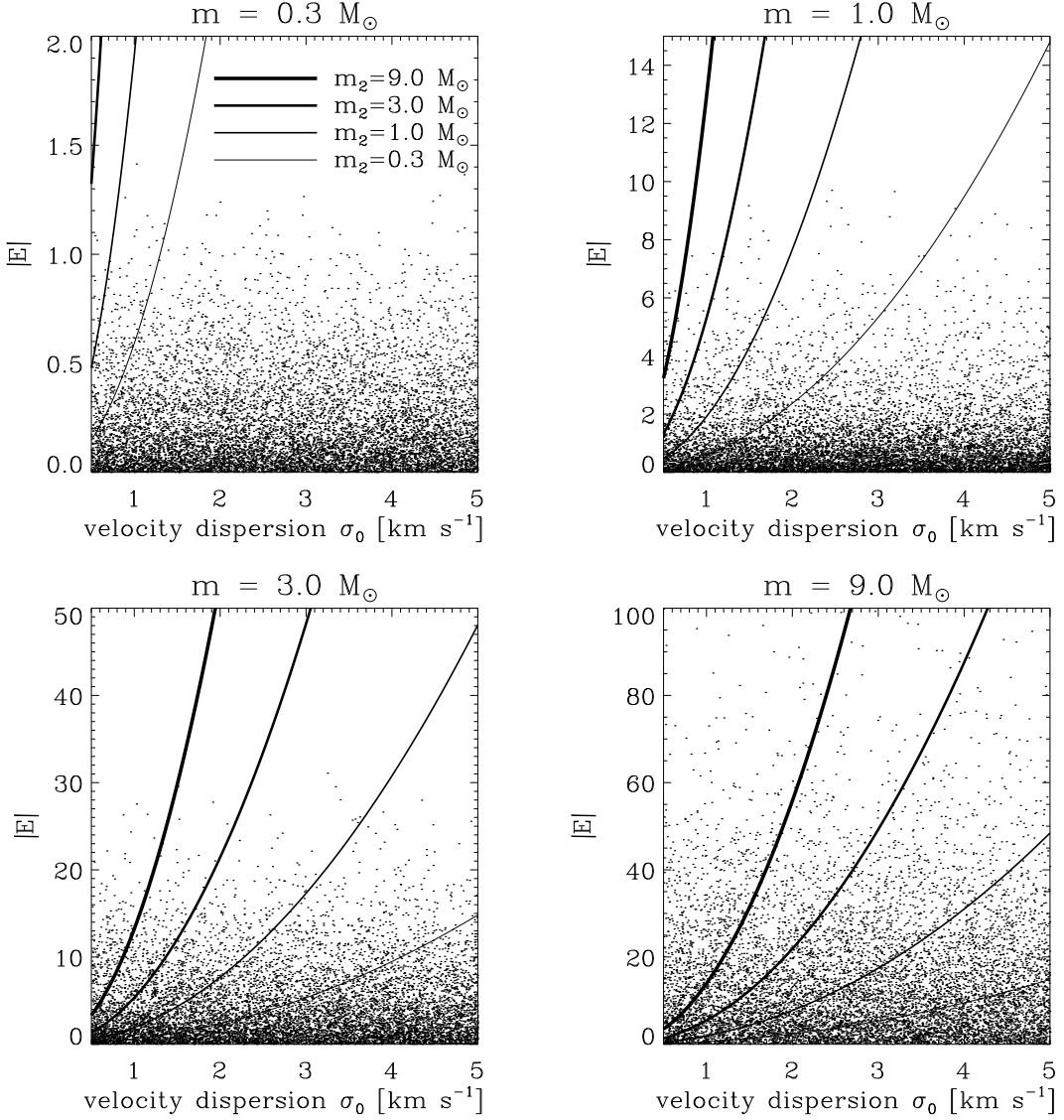}
  \caption{Energies associated with the final binary orbits ({\it points}), and encounters with other cluster members of various masses $m_{2}$ ({\it lines}).  Plot titles indicate the mass of the binary companion to the 22 \msuns primary.  The absolute value of the negative binary energy is plotted for comparison with the encounter energies.  Points above the solid lines are energetically unable to be ionized.  See section \ref{fate} for further details.}
  \label{binary_survival}
\end{figure}

Shown in figure \ref{binary_survival} are these energies for various combinations of companion mass $m$ and the mass of the second collision partner $m_{2}$.  The lines are the left hand term in equation \ref{critical_velocity} for different values of $m_{2}$, with the relative velocity equal to the most likely relative velocity, twice the velocity dispersion.  The points are the results from a Monte Carlo sampling of the simulation results.  For a random velocity dispersion, the relative velocity is drawn from a Maxwellian distribution with $\sigma = \sqrt{2} \sigma_{0}$.  The inclination angle $i$ and periastron $r_{p}$ are sampled subject to their probabilities, proportional to $sin(i)$ and $r_{p}$ respectively.  The sampling of relative velocity, inclination angle, and periastron is repeated until an encounter results in a binary.  In figure \ref{binary_survival}, the condition for a binary to be more likely to survive is that its position is above the curves, i.e. the absolute value of the binary's orbital energy is greater than the energy of the encounter with a third body. 

\begin{figure}
 \centering
  \plotone{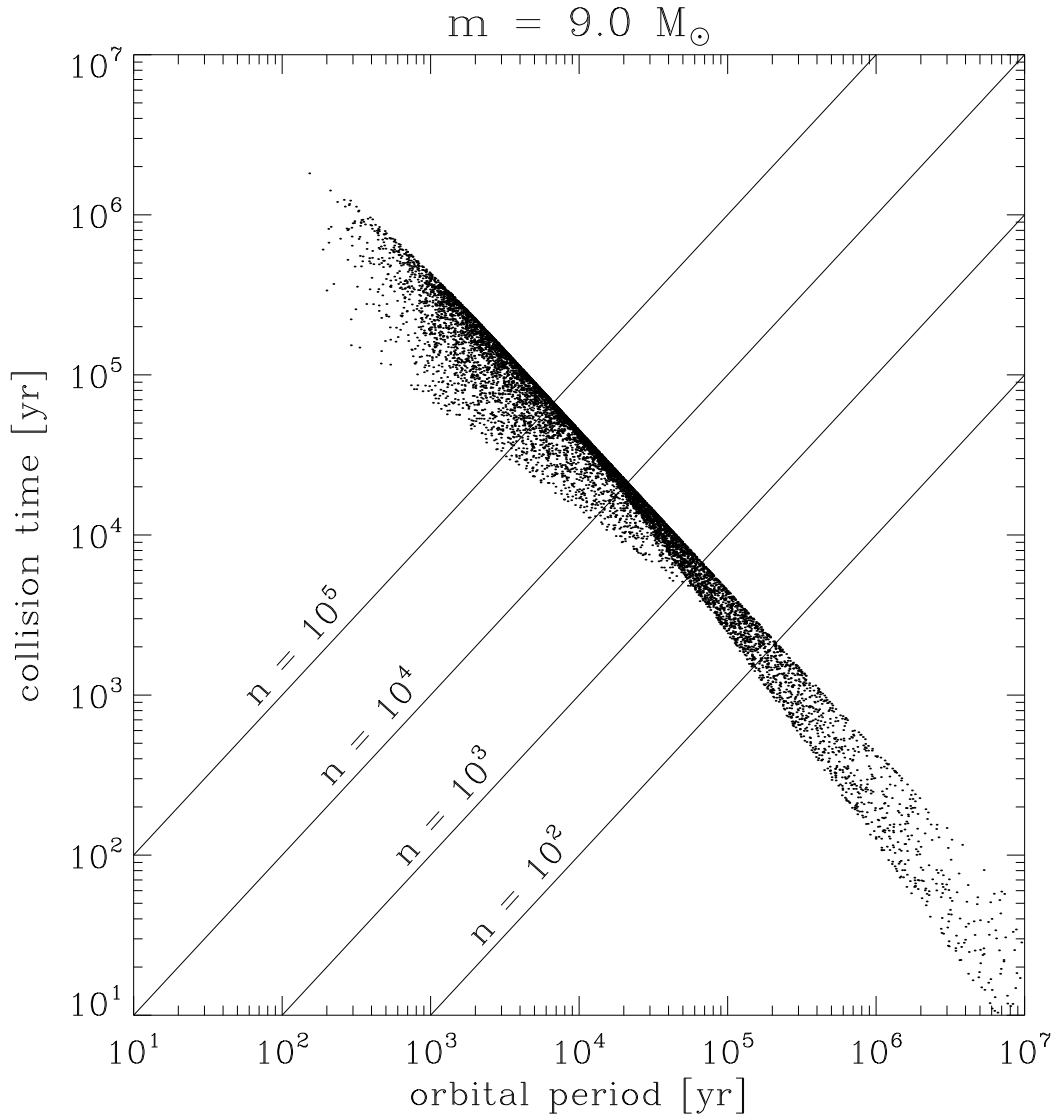}
  \caption{Encounter times versus orbital times for the same points as figure \ref{binary_survival}, for $m = 9.0$ \msun.  Points above the lines will encounter the disk a second time before the binary encounters another star in a cluster with the labeled stellar densities.}
  \label{survival_timescales}
\end{figure}

Capture formed binaries with low mass companions, $m \sim 0.3$ \msun, appear unlikely to survive for long in a dense environment.  Even encounters with slow moving, low mass cluster members are likely to ionize a typical binary.  As the mass of the captured companion increases, more of the points lie above the energy curves.  For $m > 3.0$ \msun, encounters with low mass cluster members at moderate velocities are largely unable to ionize the system, but the majority of binaries seem soft.  An effect not taken into account here is further passages through the disk.  If the orbital period of the binary is less than the encounter timescale in the cluster, the binary will harden before encounters with further cluster members.  In practice this effects only the more massive binaries; plotted in figure \ref{survival_timescales} are the orbital periods versus encounter times of the points in figure \ref{binary_survival} for $m=9.0$ \msun.  The encounter times were calculated from equation \ref{our_rate_zero}, using the average mass of a cluster member $\sim 0.5$\msun, and the semi-major axis of the binary as the encounter radius.  If the orbital period is short compared to the encounter time, the point will lie above the solid line.  The encounter time was calculated for $n = 10^{4}$ pc$^{-3}$, and adjustments for different stellar densities are plotted.

\begin{figure}
 \centering
  \plotone{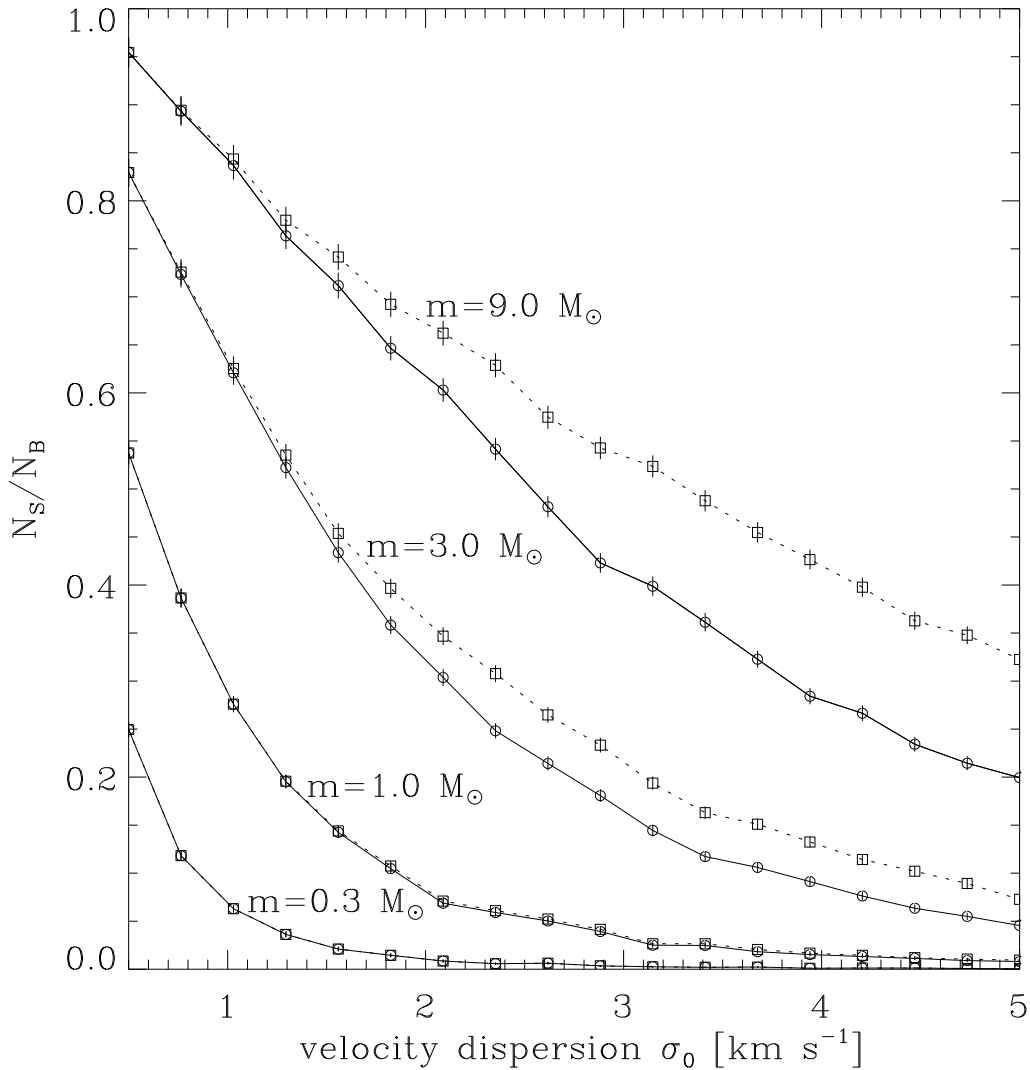}
  \caption{The number of surviving binaries $N_{S}$ after an encounter with another cluster star, divided by the number of binaries tested, $N_{B}$, based on Monte Carlo sampling of the cluster.  Solid lines show the survival percentage without accounting for second encounters, while the dotted lines show the percentage with second encounters included for $n = 2\times 10^{4}$ pc$^{-3}$.  The curves are shown for the labeled companion masses $m$.}
  \label{survival_rates}
\end{figure}

To estimate the effect of repeated disk passages, if a point lies above the line we assume that it passes through the disk a second time.  The change in energy associated with further passages is dependent on inclination angle and impactor mass \citep{moe06}; as an approximation we assume that the second passage results in one half the energy change as the first encounter.  The survival percentages of the binaries are shown in figure \ref{survival_rates}.  The figure is the result of Monte Carlo sampling of the binaries that form, and their encounters with further cluster members.  At each velocity dispersion, $4000$ binaries are created, with initial conditions chosen as detailed above.  After each binary is formed, an encounter partner is drawn from the same velocity distribution and from the IMF of equation \ref{imf}.  The energy of the binary is compared to the energy associated with the encounter.  If the encounter energy is insufficient to ionize the binary, we count the binary as a survivor.  Plotted is the number of survivors,  $N_{S}$, divided by the number of binaries tested, $N_{B}$.  Poisson counting statistics are assumed for the error bars.  Plotted for each mass are the results with and without accounting for second passages, in a stellar environment similar to the Trapezium center, $n=2\times10^{4}$ pc$^{-3}$.

When further passages through the disk are included in the calculation, there is a modest increase in the survival of high mass binaries, and essentially no change to the lower mass companions' survival rate.  The chances of a low mass ($m < 1.0$ \msun) companion surviving in a binary after further encounters with cluster siblings are low for all but the lowest velocity dispersion environments.  In a region like the Trapezium, with $\sigma_{0} ~ 2.3$ \kms, the survival percentage for 1 \msuns companions is $\sim 0.05$.  However more massive companions, with $m > 3.0$ \msun, have a much higher survival percentage.  A 9 \msuns companion cannot be ionized by $\sim$ 60 percent of further encounters in the same environment.  It seems that if stellar encounters are frequent enough for disk assisted capture to be a significant binary formation mechanism, further encounters will tend to disrupt the lower mass binaries, with higher mass binaries preferentially surviving.  An effect that this analysis does not explore is the progressive softening of a binary by repeated, non-ionizing encounters.  If an interaction results in a less tightly bound binary, the next encounter will be more likely to disrupt the system.  This is beyond the simple energetic treatment given here, but has the potential to further decrease the survival rates of binaries near the critical binding energy.  

\subsection{Observable Indications}
Binaries formed by capture would exhibit several observable characteristics.  The semi-major axis $a$ of the orbit is given by $a = G\mathcal{M}\mu/2 E$,
with $E$ the orbital energy.  Typical values for the encounters simulated here range from $a\sim 10^4$ AU for $m = 0.3$\msuns to $a\sim 10^3$ AU for $m = 9.0$\msun.  Since the periastra of the captured binaries are less than the disk radius, the eccentricities are large.  Thus the capture of an initially  unbound impactor results in a long period, highly eccentric ($e \sim 0.9$) binary.  Previous work on repeated encounters between a massive star-disk system and a less massive impactor \citep{moe06} suggests that the disk will be truncated within a few orbits to roughly the periastron distance, as the semi-major axis of the binary shrinks.  During these repeated encounters the orbit is circularized to some extent, though it remains eccentric with $0.6 \lesssim e \lesssim 0.8$ depending on the inclination angle.  The evolution of the orbit's semi-major axis is likewise inclination dependent, shrinking by anywhere from a factor of $\sim 3$ for a prograde encounter to an order of magnitude for retrograde.  In the most dissipative cases, in-plane collisions, the disk is almost totally disrupted.  In out-of-plane encounters, the disruption of the disk is more gradual, with the disk retaining material through $\sim 10$ orbits or more.  

The disruption of the circumstellar environment during capture could help explain the observed lack of disks around massive stars in the Trapezium core and other young clusters.  As opposed to young, embedded protostars, where there are examples of disks, there is no evidence of circumstellar disks around mature massive stars.  In the Trapezium, the most massive stars are multiple, and diskless.  A higher mass binary companion is more spatially disruptive to the disk during repeated passages, and is also more likely to have time to pass through several times before an encounter with another cluster star.  Regardless if the ultimate fate of the binary is survival or disruption, the initial passages through the disk provide an efficient disk disruption mechanism.  Since binary capture is destructive to the disk, the disk dissipation timescale is tied to the binary formation timescale.  In dense clusters, $n \gtrsim 4 \times 10^{4}$ pc$^{-3}$, the timescale for all massive stars to have undergone a binary capture is under $10^{6}$ yr, comparable to the length of the embedded phase, suggesting that destruction via repeated encounters could contribute to the dissipation of the disks.  This is similar to an effect noted by \citet{pfa06}, though in their scenario disk destruction around massive stars occurs via many encounters with different low mass stars, whereas here it is due to repeated encounters with a captured, moderate mass companion. 

\begin{figure}
 \centering
  \plotone{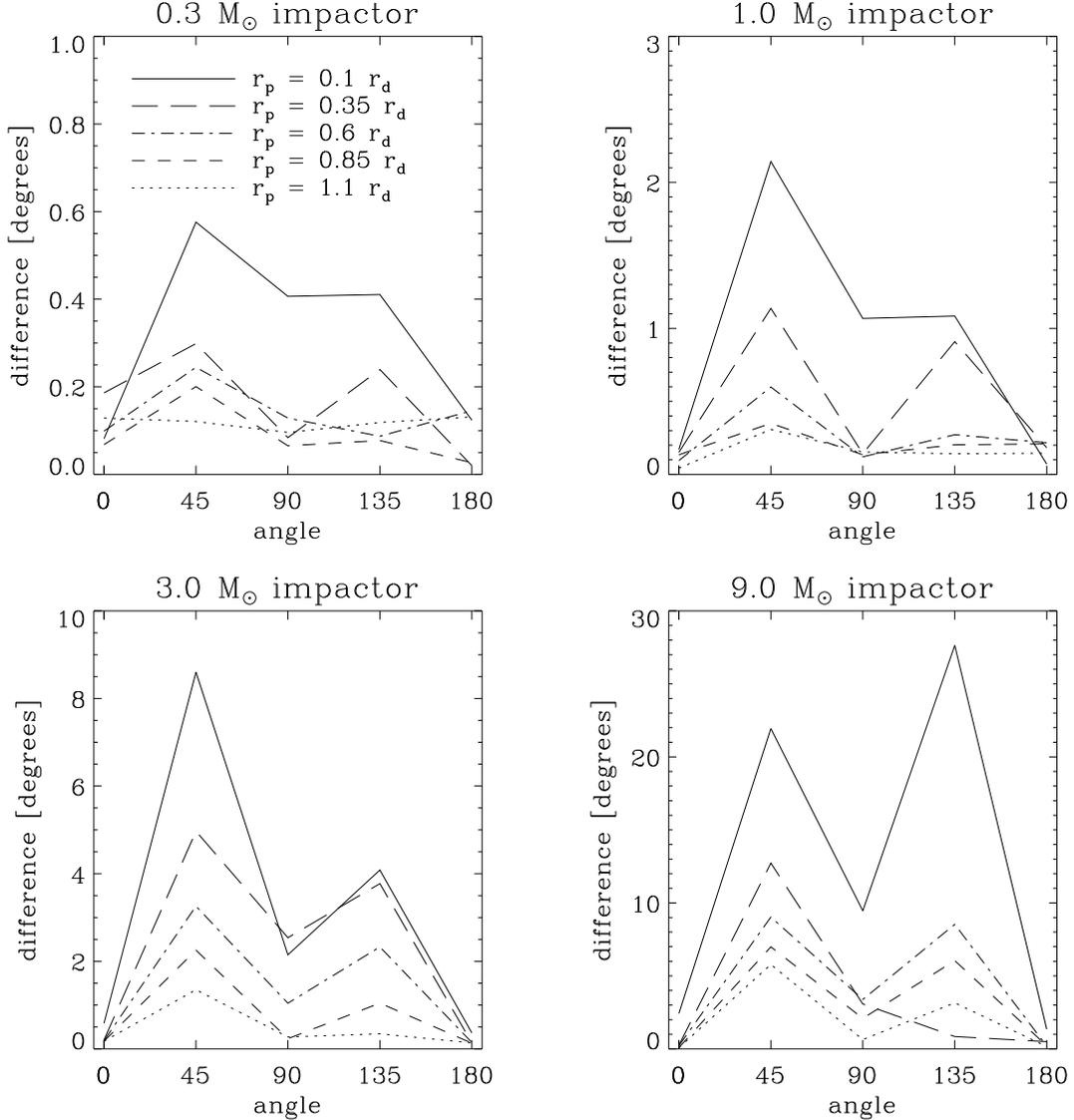}
  \caption{Change in inner disk orientation, with respect to a fixed coordinate system, for each  periastron radius $r_{p}$ and impactor mass $m$.}
  \label{disk_shift}
\end{figure}

The tilting and warping of a disk in an out-of-plane encounter has been seen in many previous simulations \citep{hel95,lar97,bof98,moe06}.  Shown in figure \ref{disk_shift} is the change in orientation, with respect to a fixed coordinate system, of the disk due to the passage of the impactors.  The same mass, inclination, and periastra as in Figure \ref{binary_vals} are plotted.  The disk orientation is calculated by averaging the angular momenta vectors of the particle orbits about the primary.  The change is given in degrees, and is simply the angle between the initial and post-passage orientation vectors, not precession about the axis of the binary orbit.  The disk is warped by these passages, so the tilt of the disk varies with radius; plotted here are the orientations measured by considering disk material within 50 AU of the primary.

The change in disk orientation from an encounter with an impactor of mass $m \lesssim 1.0$ \msuns is not likely to be more than a degree.  However, for the larger impactor masses simulated, changes in orientation of $\sim 5$ degrees or more are possible, with the most extreme encounters twisting the disk through 10s of degrees.  During out-of-plane encounters, the remnant disk experiences an almost impulsive torque during periastron passages.  Assuming that the orientation of an outflow is somewhat tied to the angular momentum of the disk that feeds it, such periodic changes could be indications of a capture \citep{bat00a,moe06}.

The statistics of mass ratios $q = m/M$ in high mass multiple systems are not constrained to the point where a comparison to this formation mechanism can be made.  Because higher mass companions preferentially survive, one would expect the distribution of mass ratios to be biased toward the higher end.  For visual binaries in Orion, \citet{pre99} concluded that the distribution of $q$ was consistent with the companions being drawn from the IMF.  In NGC6611, \citet{duc01} could neither confirm nor rule out an IMF distribution.  \citet{mas98} surveyed spectroscopic binaries among O stars, finding a flatter $q$ distribution and a lack of low $q$ systems, however these results may be partially due to bias against detecting low mass companions in spectroscopic searches.  The tendency of the binaries studied here to be relatively high $q$ seems to be currently unconstrained by observations.

\section{SUMMARY}
The capture of companions through disk interactions has been previously studied in the context of low mass stars.  In this paper we have extended the examined parameter space toward values more representative of high mass stars.  The capture rates we calculate are an order of magnitude or more higher than those found in the lower mass scenario.  Given the current stellar density and velocity parameters of a region like the Trapezium core, the rates calculated here could account for $\sim 50$\% of the massive stars having a companion after 1 Myr.  Though not high enough to account for the extremely high companion fraction of the Trapezium stars, this rate is not negligible, and the disk-capture mechanism should not be ignored in dense regions of massive star formation.  However, the binaries that form are soft and perhaps easily disrupted.

We attempted to estimate the robustness of the binaries, concluding that massive companions are more likely to survive in a dense cluster.  Lower mass companions tend to form very weakly bound binaries, which are unlikely to remain bound after encounters with other cluster members.  The same conditions that lead to a high capture rate, low velocity dispersion and high stellar density, result in a high rate of disrupting encounters.  If this mechanism is to contribute to the observed multiplicity, the most likely surviving binaries will have massive companions, and will have passed through the disk several times before another encounter occurs.   However, the details of binary-single star encounters, with one dominant mass in an eccentric binary and the remnants of a disk still around, warrant further investigation beyond the simple treatment given here.

The effects of a mass dependent velocity dispersion are not likely to affect the capture rates, except for extreme cases, close to true energy equipartition.  Mass segregation, however, enhances the capture rates significantly for more massive impactors.  In a heavily mass segregated environment, such as the Arches cluster, capture of a massive companion ($m \gtrsim 5$ \msun) would be a frequent occurrence, though in the Arches cluster itself the high velocity dispersion would lower the capture rate significantly.  Capturing a companion is destructive to the disk, and the timescale for binary capture is comparable to the expected disk lifetime in dense regions.  The likelihood of a companion being massive, and the disk destruction associated with repeated encounters, could help explain the observed lack of disks in regions like the Trapezium.

This work has ignored the effects of disk replenishment from the reservoir of gas at the center of a young cluster, and the dissipation of the disks over time.  Dissipation can occur viscously or from radiative mechanisms, and the replenishment of the disk depends sensitively on the details of the radiation field of the massive star.  Whether these combined effects enhance or decrease the rates calculated here will depend on the age of the protostars, the local gas and dust structure, and the amount of gas left in the cluster.

\vspace{.25 in}
This work was supported by NASA grant NNA04CC11A to the CU Center for Astrobiology.  We thank the referee Marc Freitag for excellent comments that improved this manuscript.  Daniel Price's program SUPERSPHPLOT was used to make the column density plots.




\end{document}